\documentclass[]{interact}

\usepackage{epstopdf}
\usepackage[caption=false]{subfig}

\usepackage[numbers,sort&compress,merge]{natbib}
\bibpunct[, ]{[}{]}{,}{n}{,}{,}

\theoremstyle{plain}

\theoremstyle{definition}

\theoremstyle{remark}

\usepackage{amsmath}
\usepackage{graphicx}
\usepackage{color}
\usepackage{amssymb}

\newcommand{\om}{\mathbf{\hat{\omega}}}

\begin{document}


\title{The effect of flexibility and bend angle on the phase diagram of hard colloidal boomerangs}

\author{
\name{Tara Drwenski\textsuperscript{a}\thanks{CONTACT Tara Drwenski. Email: t.m.drwenski@uu.nl; Ren\'e van Roij. Email: r.vanroij@uu.nl} and Ren\'e van Roij\textsuperscript{a}}
\affil{\textsuperscript{a}Institute for Theoretical Physics,  Utrecht University, Princetonplein 5, 3584 CC Utrecht, The Netherlands}
}

\maketitle

\begin{abstract}
We study the effect of flexibility and bend angle on systems of hard semiflexible boomerangs. These are modelled as two rodlike segments joined at one end with an angle that can fluctuate about a preferred angle. We use a second-virial theory for semiflexible chains with two segments, and numerically solve for the full orientation distribution function as a function of the four angles that determine the boomerang's orientation. We plot the single segment distributions as a function of two angles as well as the interarm angle distribution. For stiff boomerangs, we find prolate, oblate, and biaxial nematic phases depending on the bend angle and density, in partial agreement with previous results on rigid boomerangs. For the case that the preferred interarm angle is $90^\circ$, however, we find that the biaxial nematic phase has four-fold rather than two-fold rotational symmetry, and thus requires fourth-rank order parameters to describe it. In addition, we find that flexibility drastically reduces the region of stability for the biaxial nematic phase, with the prolate nematic becoming more favourable.
\end{abstract}


\section{Introduction}\label{sect:intro}

Steric repulsions, and therefore entropy alone, can give rise to orientationally ordered phases in systems of hard anisotropic colloids~\cite{onsager1949,deGennes1993}.  Uniaxial rodlike colloids favour the isotropic phase (I) at low densities, which maximizes their orientational entropy, and at higher densities they form a nematic phase, where they align along a director to lower their excluded volume. For needlelike rods, Onsager's second-virial theory for the isotropic-nematic transition is exact~\cite{onsager1949} and therefore forms the starting point for many extensions, including those towards less elongated or less symmetric shapes. Less symmetric rigid colloids, which need three angles to describe their orientations, can form three types of homogeneous nematic phases: an oblate nematic (N$_-$) phase where they align along their shortest axis, a prolate nematic (N$_+$) phase where they align along their longest axis, and a biaxial nematic (N$_\text{B}$) phase, where they align along both axes. This biaxial nematic phase has been long searched for in thermotropic systems, due to its potential for opto-electronic applications~\cite{luckhurst2015}. Though the theoretical prediction of the biaxial nematic goes back to the 1970s~\cite{freiser1970}, the observation of the N$_\text{B}$ phase in thermotropic systems is still disputed~\cite{luckhurst2015}. 

Common biaxial particle models for studying the existence of the N$_\text{B}$ phase are spheroplatelets, cuboids, and ellipsoids~\cite{mulder1989,straley1974,luckhurst2015}. Monte Carlo simulations have confirmed the stability of the biaxial nematic for these particle models in certain shape and density regimes. However, in order to overcome competition with spatially ordered phases, depending on the particle model, high particle aspect ratios may be necessary~\cite{allen1990,camp1997,peroukidis2013,peroukidis2013b,dussi2018}. Another biaxial particle model is that of hard boomerangs (sometimes called bent-core particles or dimers), which are usually modelled as two spherocylinders of length $L$ and diameter $D$ joined at one end with a certain interarm angle $\chi_0$. These boomerangs are not convex, and are less symmetric than the ellipsoids, spheroplatelets, or cuboids. Though ostensibly simple, boomerangs can on the basis of symmetry considerations form a large number of phases as exhaustively studied in Ref.~\cite{lubensky2002}. Recently, particles with this symmetry have received increased attention due to the fact that they can form chiral phases despite being achiral themselves~\cite{takezoe2006,dozov2001,greco2015}. 

Hard needlelike boomerangs are predicted to have a so-called Landau point with a direct I-N$_\text{B}$ transition for opening angles $\chi_0=107^\circ$, with the boomerangs preferring prolate ordering above this angle and oblate ordering below this angle~\cite{teixeira1998}. Thermotropic systems have a similar predicted Landau angle~\cite{luckhurst2001}. For lower aspect ratio boomerangs, the Landau point has been shown to shift to smaller opening angles within second-virial theory~\cite{teixeira1998}, whereas third-virial calculations have been shown to increase the Landau angle~\cite{camp1999}, but in both of these cases the phase diagram topology is unaffected. We expect that second-virial theory is exact in the limit that $L/D\to\infty$~\cite{onsager1949}, even for boomerangs, and in this limit the isotropic-nematic transition occurs at such a low density that competition with positionally ordered phases is not to be expected. Simulations of hard boomerangs, however, have yet to confirm this phase diagram topology~\cite{camp1999,lansac2003,dewar2004,dewar2005}. One possible explanation for the yet unobserved biaxial nematic phase of hard boomerangs is that so far only boomerangs with relatively low aspect ratios have been simulated, where spatially ordered phases may have preempted the N$_\text{B}$ as well as the N$_-$ phase. In addition, these simulations have suffered from long equilibration times as these systems tend to jam close to the Landau point.

Many colloidal rods with high aspect ratios are not actually rigid particles, but semiflexible~\cite{vroege1992, purdy2005}. This semiflexibility has been shown to be key in describing the phase behaviour of binary mixtures of fd virus, as the flexibility changes the effective aspect ratio of the fd virus depending on the state point~\cite{dennison2011PRL,dennison2011JCP}. This was done using a theory originally developed by Wessels and Mulder~\cite{wessels2003,wessels2006}, which is based on a simple model of semiflexibility that relies on discretising a non-convex rodlike particle into a chain of connected rigid rod segments with a bending potential that gives rise to the stiffness of the particle. 

Interestingly, a recent work~\cite{vaghela2017} extended the analysis of Ref.~\cite{teixeira1998} from rigid boomerangs to flexible ones, and concluded that flexible boomerangs with a preferred straight configuration ($\chi_0=180^\circ$) can form a biaxial nematic phase. The approach of Ref.~\cite{vaghela2017} is a second-virial theory with a segmentwise approximation for the excluded volume of two boomerangs and with an interarm bending potential. In addition, the second-virial theory is further simplified using the method of Straley~\cite{straley1974}, by considering the excluded volume of six discrete orientations of two particles and interpolating between these, which allows for all angular dependencies to be written in terms of a basis of second-rank Wigner rotation matrices.

Our purpose here is to investigate the effect of flexibility on the stability of the biaxial nematic phase for boomerangs with various preferred angles, using the method of Refs.~\cite{wessels2003,wessels2006}. We expect our approach to be similar to that of Ref.~\cite{teixeira1998} for very stiff boomerangs, since we also use a second-virial theory with a segmentwise approximation for the excluded volume. One important difference, however, is that we solve for the complete orientation distribution functions, instead of only the second-rank order parameters. In addition, we use the full form for the excluded volume within the segmentwise approximation, without interpolation or other simplifications.

 This paper proceeds as follows. In Sec.~\ref{sect:method}, we explain the second-virial density functional theory for hard semiflexible chains in the case of boomerangs, that is, two-segment chains, based upon Refs.~\cite{wessels2003,wessels2006}. Additionally, we give the order parameters used to distinguish between the different phases. In Sec.~\ref{sect:results}, we give our results for stiff and flexible boomerangs with various preferred opening angles. Finally, we discuss these results and conclude in Sec.~\ref{sect:conclusions}.

\section{Method}\label{sect:method}

\begin{figure}[tbph]
	\centering
			\includegraphics[width=0.8\textwidth]{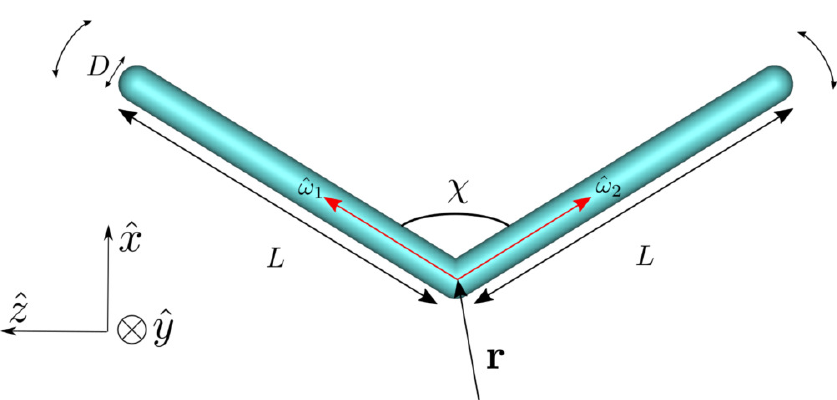}
		\caption{Our model of a flexible boomerang at position $\mathbf{r}$ consisting of two spherocylinders joined at one end with an interarm angle of $\chi$, arm lengths $L$ and diameter $D$. The red arrows show the orientations of the individual segments $\om_1$ and $\om_2$. Shown is the case with $\chi=117^\circ$.}\label{fig:particleModel}
	\end{figure}

We use the formalism of Wessels and Mulder~\cite{wessels2003,wessels2006} for semiflexible chains, but only consider boomerangs, that is, chains with only two segments, where each segment is a spherocylinder of length $L$, diameter $D$, and $L \gg D$. A configuration of such a boomerang can be given by $\Omega=(\mathbf{\hat{\omega}}_1,\mathbf{\hat{\omega}}_2)$ where $\mathbf{\hat{\omega}}_m$ is the unit vector describing the orientation of the $m$th uniaxial segment ($m=1,2$). We also introduce the planar interarm angle $\chi$ (see Fig.~\ref{fig:particleModel}), defined by $\cos \chi = \om_1 \cdot \om_2$.

In density functional theory, we express the free energy as a functional of the single-particle density $\rho(\mathbf{r},\Omega)$. We assume that the single-particle density has no spatial dependence, i.e. $\rho(\mathbf{r},\Omega) = \rho \psi(\Omega)$, where $\rho = N/V$ is the average density in a system of $N$ particles and volume $V$, and $\psi(\Omega)d\Omega$ is the probability to find a particle with orientation $\Omega$ in the infinitesimal interval $d\Omega$. The free energy per particle in the second-virial approximation can be written as
	\begin{eqnarray}\label{eq:freeEnergy}
	    \frac{\beta F\left[\psi(\Omega)\right]}{N} &=& 
	        \ln\mathcal{V}\rho-1+\int d\Omega\,\psi(\Omega)\left[\ln\psi(\Omega)  +u(\Omega)\right]\nonumber\\
	         &+& \frac{\rho }{2}\, \int d\Omega \int d\Omega' E(\Omega,\Omega')\psi(\Omega)\psi(\Omega) ,
	\end{eqnarray}
 where $\mathcal{V}$ is the thermal volume, $\beta = 1/(k_B T)$ is the inverse thermal energy, and $k_B T u(\Omega)$ is a configuration dependent bending energy of a single boomerang, which is an irrelevant constant for a rigid boomerang with a fixed opening angle $\chi$ but is of physical significance for flexible rods as we will see below. The excluded volume $E(\Omega,\Omega')$ [in Eq.~\eqref{eq:freeEnergy}] between two boomerangs with orientations $\Omega$ and $\Omega'$ is defined as
	\begin{eqnarray}\label{eq:exclVol2}
	    E(\Omega,\Omega') &=& -\int d\mathbf{r}_{12} \, f(\mathbf{r}_{12},\Omega,\Omega') \nonumber\\
	    &=& -\int d\mathbf{r}_{12}\, (\exp\left[-\beta U(\mathbf{r}_{12},\Omega,\Omega')\right] - 1),
	\end{eqnarray}
where $f(\mathbf{r}_{12},\Omega,\Omega')$ is the Mayer function, $U(\mathbf{r}_{12},\Omega,\Omega')$ is the pair potential, and $\mathbf{r}_{12} = \mathbf{r} -\mathbf{r}'$ is the vector connecting the centres of the two particles. For hard particles we assume the pair potential to be
	\begin{equation}
	\label{eq:potential}
	 \beta  U(\mathbf{r}_{12},\Omega,\Omega')= \left\{
	     \begin{array}{cl}
	       \infty, & \text{1 and 2 overlap;}  \\
	       0, & \text{otherwise} .
	     \end{array}
	   \right. 
	\end{equation}
However, since the excluded volume [Eq.~\eqref{eq:exclVol2}] of two chains is complicated to calculate for all configurations, we follow Ref.~\cite{wessels2003,wessels2006} and use a segmentwise approximation for the excluded volume, i.e., we write
	\begin{equation}\label{eq:exclVolSeg}
		E(\Omega,\Omega') = \sum_{m,m'=1}^2 e(\om_m,\om'_{m'}),
	\end{equation}
where $e(\om_m,\om'_{m'})$ is simply the excluded volume between two needlelike segments with $L\gg D$, given by~\cite{onsager1949}
	\begin{equation}\label{eq:exclVol}
		e(\om,\om') = 2L^2D \sqrt{1 - (\om \cdot \om')^2}.
	\end{equation}
This segmentwise approximation [Eq.~\eqref{eq:exclVolSeg}] neglects the polarity of the bent boomerangs, and it overestimates the true excluded volume worse for smaller interarm angles $\chi$~\cite{bisi2008}. 

We assume that the boomerangs have a bending energy between their segments, given by~\cite{wessels2003,wessels2006}
	\begin{equation}\label{eq:bendPot}
		u(\om_1,\om_2) = -\frac{P}{L} \cos \left[\chi(\om_1 \cdot \om_{2})-\chi_0 \right],
	\end{equation}
where $P$ is the persistence length and $\chi_0$ is the preferred configuration of the boomerang, e.g., when $\chi_0=180^\circ$ the boomerang fluctuates around a straight rod configuration.

Equation~\eqref{eq:freeEnergy} can be minimized with respect to the orientation distribution function (ODF) $\psi(\Omega)$ under the normalization constraint $\int d\Omega \psi(\Omega) = 1$, the minimizing $\psi(\Omega)$ being the equilibrium ODF~\cite{vroege1992}. The resulting Euler-Lagrange equation can be written as the non-linear self-consistency equation
	\begin{equation}\label{eq:EL1}
		\psi(\Omega) = \frac{1}{Z} \exp\left[-u(\Omega)-V(\Omega) \right],
	\end{equation}
	where $Z=\int d\Omega \exp[-u(\Omega)-V(\Omega)]$ ensures the normalization of $\psi$ and we define the self-consistent field $V$ as
	\begin{equation}
		V(\Omega) = \rho \int d\Omega' E(\Omega,\Omega') \psi(\Omega').
	\end{equation}

Although one could in principle solve Eq.~\eqref{eq:EL1} on a four-dimensional grid of polar and azimuthal angles of the two segments, it is more practical to introduce single segment distributions $\psi_m(\om_m)$ that only depend on a single pair of polar and azimuthal angles of the segment $m=1,2$.
 This is done by projecting the full ODF of the boomerang $\psi(\Omega)$ onto a given segment as~\cite{wessels2003,dennison2011JCP}
	\begin{equation}\label{eq:segODF}
		\psi_m(\om_m) = \int d\hat{\omega}_{\bar{m}} \, \psi(\Omega),
	\end{equation}
where $\bar{m}\neq m$ is the remaining segment. Inserting Eq.~\eqref{eq:EL1} into Eq.~\eqref{eq:segODF} results in the set of self-consistency equations for $m=1,2$~\cite{wessels2003,wessels2006}
	\begin{equation}\label{eq:ELseg}
		\psi_m(\om_m) = \frac{1}{Z} q_1(\om_1) \exp \left[ -v(\om_m) \right] q_2(\om_2),
	\end{equation}
where the partial-chain partition functions are given by
	\begin{equation}\label{eq:partFn}
		q_2(\om_2) = \int d\om_{1} q_1(\om_1) \exp\left[ -v(\om_1)-u(\om_1,\om_2) \right],
	\end{equation}
and using the normalization of $\psi_m$ we choose $q_1(\om_1) = 1$. Here the single-segment self-consistent field is defined as
	\begin{equation}\label{eq:vField}
		v(\om_m) = \rho \sum_{m'=1}^2 \int d\om_{m'}' e(\om_m,\om_{m'}') \psi_{m'}(\om_{m'}'),
	\end{equation}
such that $V(\Omega) = \sum_{m=1}^2 v(\om_m)$.

We choose a coordinate system ($X$, $Y$, $Z$) where a segment's orientation is given by $\om_m = (\sin\theta_m \cos \phi_m,\sin\theta_m \sin \phi_m,\cos\theta_m)$ with $\phi_m$ the azimuthal angle and $\theta_m$ the polar angle with respect to $\hat{Z}$. We then numerically solve Eqs.~\eqref{eq:ELseg}-\eqref{eq:vField} using an iterative scheme for $\psi_1$ and $\psi_2$ on a discrete grid of $\theta$ and $\phi$ angles, using a uniform grid of $N_\theta=60$ and $N_\phi=60$ angles~\cite{herzfeld1984,vanRoij2005}. We expect this to be sufficiently accurate based on similar calculations for uniaxial rods, which showed that $N_\theta=40$ gave an order parameter with two digits of accuracy at the isotropic-nematic coexistence~\cite{vanRoij2005}.

Note that we find perfect symmetry between the two segments, due to symmetry in the excluded volume and bending potentials [Eqs.~\eqref{eq:exclVolSeg} and \eqref{eq:bendPot}], and so we always find that $\psi_1(\om) =\psi_2(\om)$. From the equilibrium ODFs $\psi_m(\om_m)$, we can compute the free energy, pressure, and chemical potential~\cite{wessels2003,wessels2006}, and hence the phase diagram following the standard procedure. For convenience we define a dimensionless density $c = \rho(\pi/4) (2L)^2 D $, which reduces to the usual definition for the case of a straight rigid rod of length $2L$ and diameter $D$.

In addition, we recover the full equilibrium ODF from the segment ODFs using~\cite{wessels2003}
	\begin{equation}
		\psi(\om_1,\om_2) = \frac{1}{Z} \exp \left[ -v(\om_1)-v(\om_2) -u(\om_1,\om_2)\right] .
	\end{equation}
It also turns out to be convenient to introduce (i) the boomerang's frame ($x$, $y$, $z$), defined to be the orthogonal basis proportional to ($\om_1+\om_2$, $\om_1 \times \om_2$, $\om_1-\om_2$), see also Fig.~\ref{fig:particleModel}, and (ii) the Euler angles $\alpha,\beta,\gamma$ that transform the particle frame to a reference frame, which together with the interarm angle $\chi$ fully determine the particle configuration (and are equivalent to $\om_1,\om_2$). We define the orientational average as $\langle \cdot \rangle = \int d\om_1 \int d\om_2 (\cdot) \psi(\om_1,\om_2)$. This allows us to calculate the probability density $g(\chi)$ for an internal configuration with interarm angle $\chi$, which we define as 
\begin{equation}
	g(\chi) =  \langle \delta(\chi-\arccos(\om_1 \cdot \om_2)) \rangle,
\end{equation}
where $\delta$ is the Dirac delta function. We define the average interarm angle as $\langle \chi \rangle$ and the standard deviation of the bending fluctuations as
	\begin{equation}\label{eq:OrderParSigma}
		\sigma_\chi = \sqrt{\langle \chi^2 \rangle - \langle \chi \rangle^2}.
	\end{equation}

In order to be able to distinguish and characterize the (symmetries of the) equilibrium ODF, we also define four order parameters following the notation of Rosso~\cite{rosso2007} 
	\begin{eqnarray}
		S &=& \frac{1}{2} \langle (3 \cos^2 \beta -1) \rangle \label{eq:OrderParS},\\
		U &=& \frac{\sqrt{3}}{2} \langle \sin^2\beta \cos 2\gamma \rangle \label{eq:OrderParU},\\ 
		P &=& \frac{\sqrt{3}}{2} \langle \sin^2 \beta \cos 2\alpha \rangle \label{eq:OrderParP},\\ 
		F &=&  \langle \frac{1}{2}(1+\cos^2 \beta) \cos2\alpha \cos2\gamma \,- \cos\beta \sin2\alpha \sin2\gamma \rangle.\label{eq:OrderParF}
	\end{eqnarray}
In the isotropic phase (I) all four of these order parameters are zero. A uniaxial nematic phase has $S$ nonzero and $P=F=0$, where $S>0$ corresponds to a prolate nematic phase N$_+$ and $S<0$ to an oblate nematic phase N$_-$. Note that $U \neq 0$ if the particles are biaxial as we have here for $\chi_0 \neq 180^\circ$. In a biaxial nematic phase (N$_\text{B}$), all four are nonzero with $P$ describing the phase biaxiality, and $F$ describing both the phase and particle biaxiality. We also consider the segment order parameters $S_m = \frac{1}{2} \langle (3 \cos^2 \theta_m -1) \rangle$, where $\theta_m$ is the polar angle with respect to the nematic director $\hat{n}$, which we determine as the eigenvector with the largest eigenvalue ($S_m$) of the diagonalized segment ordering tensor~\cite{deGennes1993}.

For the case of a boomerang with a preferred angle of $\chi_0= \pi/2$, we do not expect biaxial order with a two-fold rotational symmetry, but instead four-fold rotational symmetry (called the D$_4$ phase in Ref.~\cite{blaak1998}), and so we also define the additional fourth-rank order parameter~\cite{blaak1998}
	\begin{equation}\label{eq:OrderParC}
	 	C = \cos^8 \frac{\beta}{2} \cos[4(\alpha+\gamma)] + \sin^8 \frac{\beta}{2} \cos[4(\alpha-\gamma)].
	 \end{equation} 
In the isotropic or uniaxial nematic phase $C=0$, while for an N$_\text{B}$ phase $F\neq0$ and $C\neq0$ and in the D$_4$ phase $F=0$ and $C\neq0$~\cite{blaak1998}.

Due to our discrete grid of $\theta$ and $\phi$ angles, the Euler angles will sometimes not be correctly distributed (e.g. $\gamma$ is not even defined in the case of straight rods), and so we will set a threshold of $0.1$ for the absolute value of nonvanishing order parameters.

\section{Results}\label{sect:results}

\begin{figure*}[tbph]
	\centering
			\includegraphics[width=\textwidth]{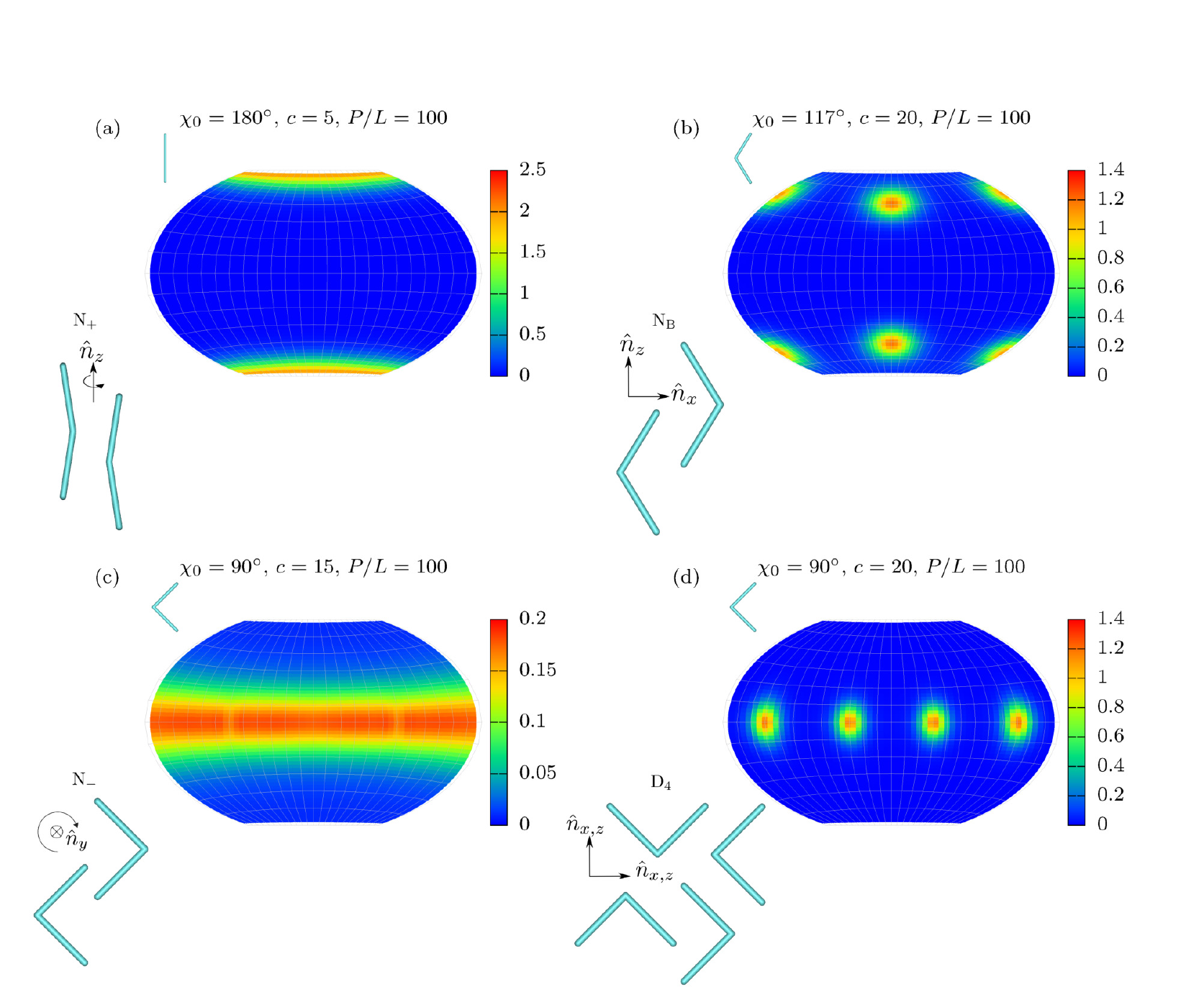}
		\caption{Examples of segment orientation distribution functions $\psi_1(\theta,\phi)$ for stiff particles with $P/L=100$ for various preferred angles $\chi_0$ and densities $c$. For $\chi_0=180^\circ$ and $c=5$ (a) we find a prolate nematic N$_+$. For $\chi_0=117^\circ$ and $c=20$ (b) we find a biaxial nematic N$_\text{B}$ where boomerangs align their long axis $\hat{z}$ with the pole. For $\chi_0=90^\circ$ and $c=15$ (c) we find a oblate nematic N$_-$ with director parallel to the pole. For $\chi_0=90^\circ$ and $c=20$ (d) we find a D$_4$ phase with boomerangs having four equivalent preferred orientations related by a rotation of $\pi/2$. Illustrations in the upper left corners show a boomerang with the corresponding interarm angle $\chi_0$. Illustrations in the lower left corner show a schematic representation of each phase with the subscript on the nematic director $\hat{n}$ indicating which particle axis is aligned along it, and with arrows around the director indicating symmetry under rotations around the director.}\label{fig:psi}
	\end{figure*}

We first consider stiff particles with a persistence length of $P/L=100$, which corresponds to bending fluctuations on the order $\sigma_\chi \lesssim 6^\circ$, with these fluctuations only weakly depending on density. The single-segment ODF, together with information about the interarm angle can provide a qualitative understanding of the full boomerang ODF, which is a function of four angles. Therefore, in Fig.~\ref{fig:psi}, we show the equilibrium single-segment ODF $\psi_1(\theta,\phi)$ on the grid of the $\theta$ and $\phi$ angles using the Winkel Tripel map projection for ease of viewing, for various densities $c$ and preferred opening angles $\chi_0$. We also include a schematic representation of the possible phases in the lower left corner of each plot. In Fig.~\ref{fig:psi}(a), the boomerangs prefer a straight orientation ($\chi_0=180^\circ$), and at density $c=5$ we find a prolate nematic phase where segments prefer orientations parallel or antiparallel to the nematic director $\hat{n}$ along the map pole. They also prefer to be essentially antiparallel to each other, since $\sigma_\chi \approx 3.1^\circ$ and $\langle \chi \rangle \approx 174^\circ$. Next, in Fig.~\ref{fig:psi}(b), we consider particles with an intrinsic biaxiality due to a preferred opening angle $\chi_0=117^\circ$, which at density $c=20$ form a biaxial nematic phase. Here we find the average interarm angle to be $\langle \chi \rangle \approx 119^\circ$ and the standard deviation to be $\sigma_\chi \approx 5.5^\circ$. We conclude that if the first segment has an orientation e.g. in the peak in the upper left of Fig.~\ref{fig:psi}(b), then the second segment must have an orientation approximately given by the peak in the lower left, or else the particle's interarm angle would differ significantly from the average interarm angle. Therefore, in this N$_\text{B}$ phase, particles have two preferred orientations related by the transformation $\hat{x} \to -\hat{x}$ and so the segment ODF has four peaks. For a preferred angle of $\chi_0=90^\circ$, the particles are platelike and stiff with $\langle \chi \rangle \approx 90^\circ$ and $\sigma_\chi \approx 5.6^\circ$. As evident from the single equatorial peak in Fig.~\ref{fig:psi}(c) for $c=15$, we find that they form an oblate nematic with $\hat{n}$ along the pole. Finally, in Fig.~\ref{fig:psi}(d) we see that for $\chi_0=90^\circ$ and $c=20$, the boomerangs form a D$_4$ phase with four-fold symmetry, with the four preferred orientations being related by the transformations $\hat{x} \to -\hat{x}$, $\hat{x} \to \hat{z}$, and $\hat{x} \to -\hat{z}$.

\begin{figure*}[tbph]
	\centering
			\includegraphics[width=\textwidth]{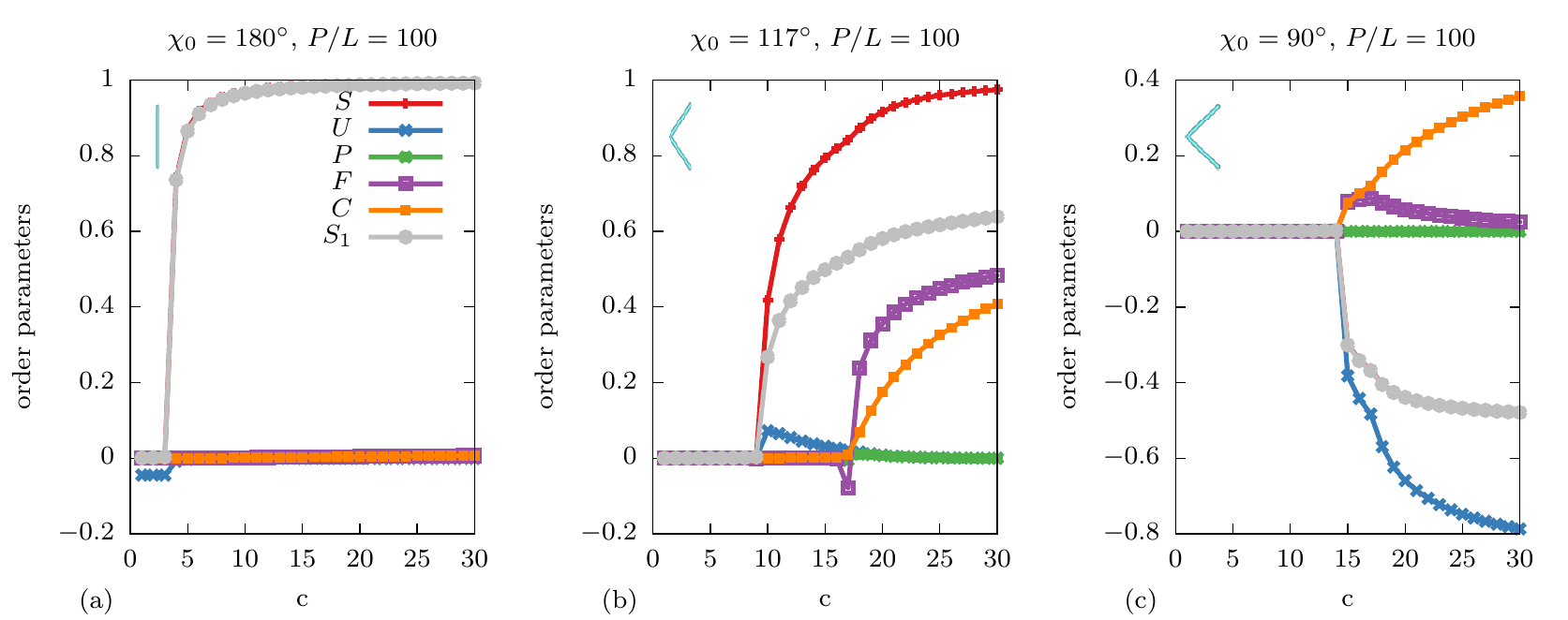}
		\caption{Order parameters defined in Eqs.~\eqref{eq:OrderParS}-\eqref{eq:OrderParC} as a function of density $c$ for stiff boomerangs ($P/L=100$) with preferred angles (a) $\chi_0=180^\circ$, (b) $\chi_0=117^\circ$, and (c) $\chi_0=90^\circ$. The key applies to (a)-(c).}\label{fig:orderPar100}
	\end{figure*}

After this illustration of the nature of the single-segment distributions $\psi_1(\om)$, we now use the full ODF $\psi(\om_1,\om_2)$ to calculate the order parameters defined from the particle frame [Eqs.~\eqref{eq:OrderParS}-\eqref{eq:OrderParC}]. These are shown in Fig.~\ref{fig:orderPar100} as a function of the density $c$ for stiff boomerangs ($P/L=100$) with preferred opening angles of (a) $\chi_0=180^\circ$, (b) $\chi_0=117^\circ$, and (c) $\chi_0=90^\circ$. For the rodlike particles in Fig.~\ref{fig:orderPar100}(a), we find the expected first order I-N$_+$ transition with coexisting isotropic density $c_i \approx 3.34$ and nematic density $c_n \approx 4.17$, which we determine using the conditions of mechanical and chemical equilibrium, and which are very similar to those of rigid uniaxial rods. We note that the fact that $U$ is a small nonzero number at low densities is an artifact of calculating the Euler angle $\gamma$ for a particle with segments restricted to our numerical grid, and is not physically meaningful. Also, we note that the segment order parameter $S_1\approx S$ since $S$ measures alignment of the particle's $\hat{z}$ axis (see Fig.~\ref{fig:particleModel}), which in this case is approximately the same as the segment orientation. In Fig.~\ref{fig:orderPar100}(b), we find an I-N$_+$ transition at $c_i\approx9.55$ and $c_n\approx9.70$ followed by an N$_+$-N$_\text{B}$ transition at $c\approx 18$ which we determine by comparing the absolute value of the biaxial order parameters $P$ and $F$ to the threshold of 0.1. Since $S>S_1$ in this case, the main particle axis $\hat{z}$ is more aligned with the nematic director at high density than the segments are due to the bent shape of the particle. Finally, in Fig.~\ref{fig:orderPar100}(c), we find a very weakly first order I-N$_-$ transition at $c_i\approx c_n \approx 14$ and an N$_-$-D$_4$ transition at $c\approx 16$.

\begin{figure*}[tbph]
	\centering
			\includegraphics[width=\textwidth]{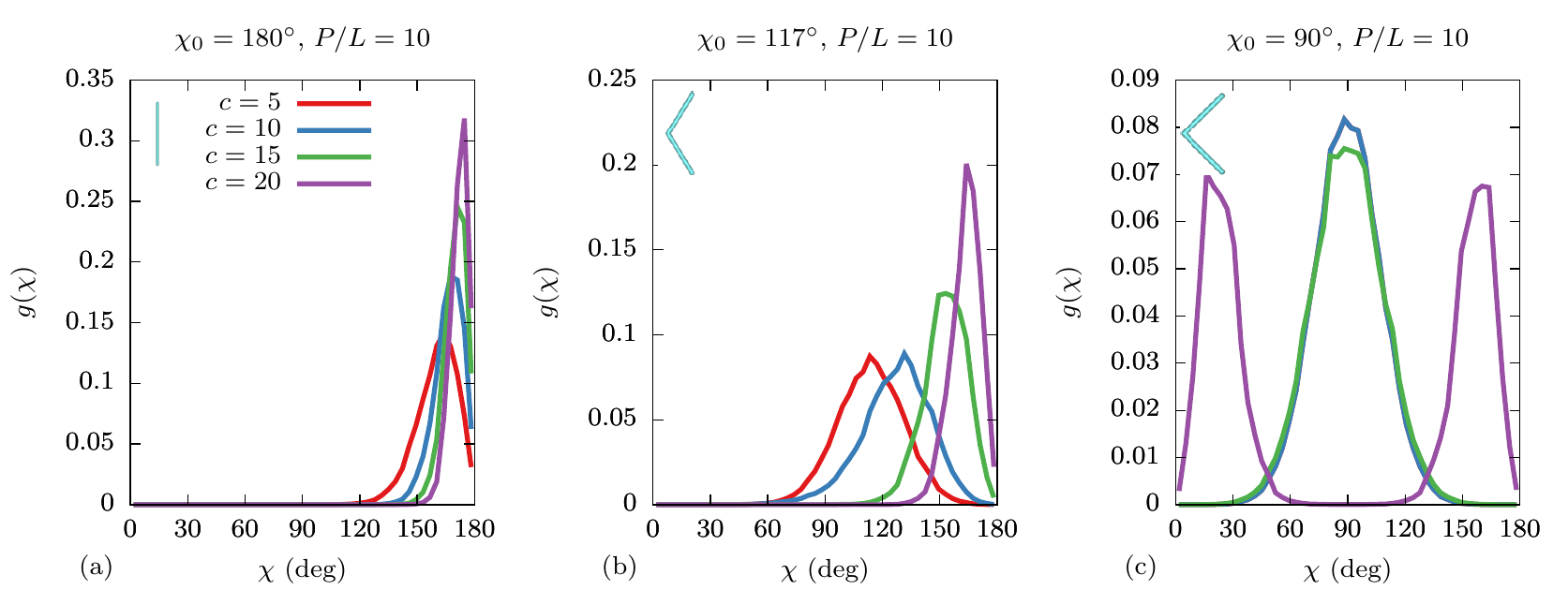}
		\caption{Probability density $g$ of the interarm angle $\chi$ for semiflexible boomerangs ($P/L=10$) at densities $c=5,10,15,20$ for preferred angles (a) $\chi_0=180^\circ$, (b) $\chi_0=117^\circ$, and (c) $\chi_0=90^\circ$. In (a), all four densities shown correspond to N$_+$ phases. In (b), $c=5$ corresponds to the isotropic phase, while $c=10,15,20$ correspond to the N$_+$ phase. In (c), $c=5,10$ correspond to the isotropic phase (we note that the blue and red curves are on top of each other), while $c=15$ corresponds to the N$_-$ phase and $c=20$ corresponds to the N$_+$ phase. The key applies to (a)-(c).}\label{fig:chiDist}
	\end{figure*}

Next we consider semiflexible boomerangs with $P/L=10$. In this case the bending fluctuations have a greater dependence on density, and so in Fig.~\ref{fig:chiDist} we plot the interarm probability density $g(\chi)$ for several densities $c$ and for three preferred angles (a) $\chi_0=180^\circ$, (b) $\chi_0=117^\circ$, and (c) $\chi_0=90^\circ$. We see in Fig.~\ref{fig:chiDist}(a) that this distribution becomes more peaked and shifts to higher $\chi$ with increasing $c$. This effect is more pronounced in Fig.~\ref{fig:chiDist}(b), where the boomerangs have $\langle \chi \rangle \approx \chi_0$ at low densities, but pay a bending energy to straighten and hence to pack more efficiently at higher densities. In Fig.~\ref{fig:chiDist}(c), we see that at densities $c \leq 15$ the particles fluctuate around $\langle \chi \rangle \approx \chi_0 = 90^\circ$, but at high density $c=20$, $g(\chi)$ has two peaks, one at small $\chi$ where segments are almost bent on top of each other and one at large $\chi$ where the particles are roughly straight. This is an artifact of our segmentwise excluded volume approximation, in which these two configurations have the same excluded volume and  also cost the same bending energy because $\chi_0=90^\circ$. In this case the full excluded volume as well as intersegment excluded volume should actually be considered. We will use the small-$\chi$ peaks that may develop in $g(\chi)$ to inform us of the break down of our model at high densities and high flexibilities.

\begin{figure*}[tbph]
	\centering
			\includegraphics[width=\textwidth]{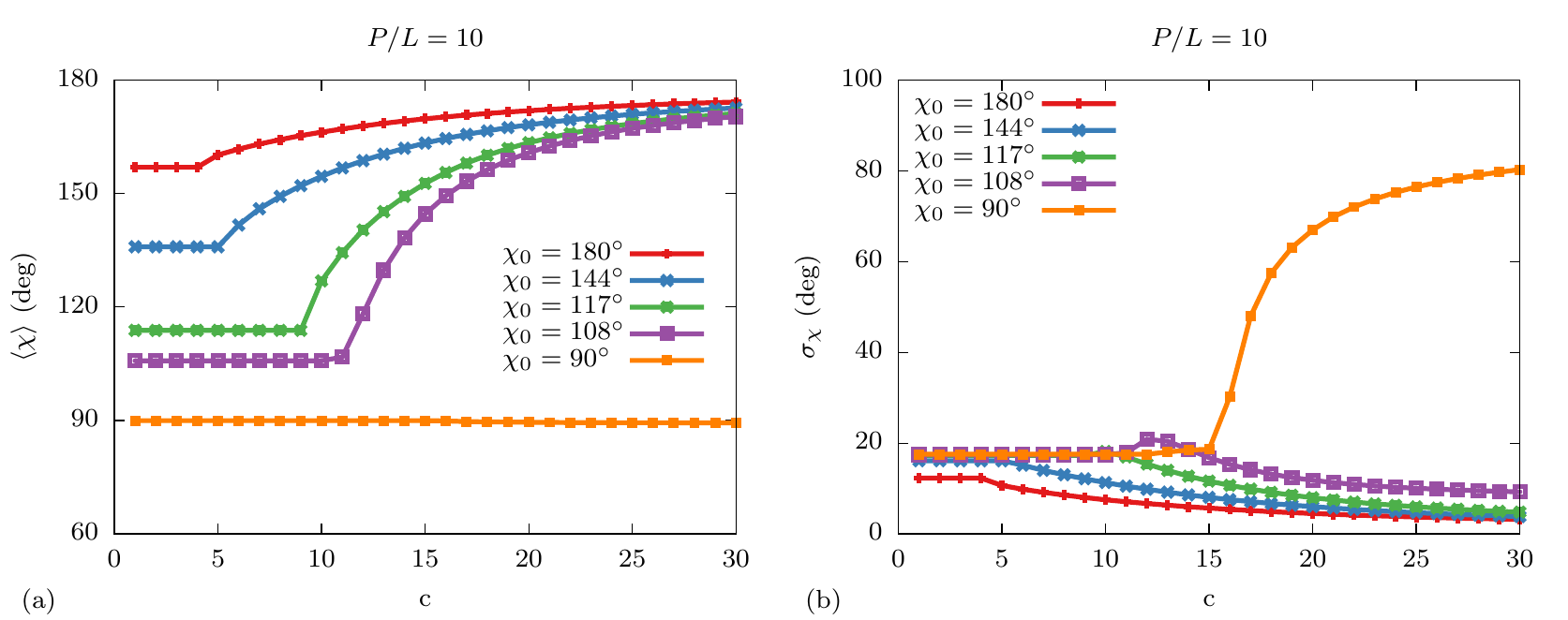}
		\caption{(a) Average interarm angle $\langle \chi \rangle$ and (b) standard deviation of the interarm angle $\sigma_\chi$, both as a function of the density $c$ for flexible boomerangs with $P/L=10$ and various preferred angles $\chi_0$.}\label{fig:chiVar}
	\end{figure*}

In Fig.~\ref{fig:chiVar}, we plot the average interarm angle $\langle \chi \rangle$ in (a) and the standard deviation $\sigma_\chi$ in (b), both as a function of the density $c$ for five different preferred angles $\chi_0$. As discussed, in the case of $\chi_0=90^\circ$, our approximation breaks down at $c>15$ where $\sigma_\chi$ becomes exceedingly large due to the spurious small-$\chi$ peak that develops. In all other cases, however, the particles tend to straighten with increasing density ($\langle \chi \rangle$ approaches $180^\circ$), which costs bending energy but reduces their excluded volume. In addition, they tend to fluctuate less with increasing density ($\sigma_\chi$ decreases).

\begin{figure*}[tbph]
	\centering
			\includegraphics[width=\textwidth]{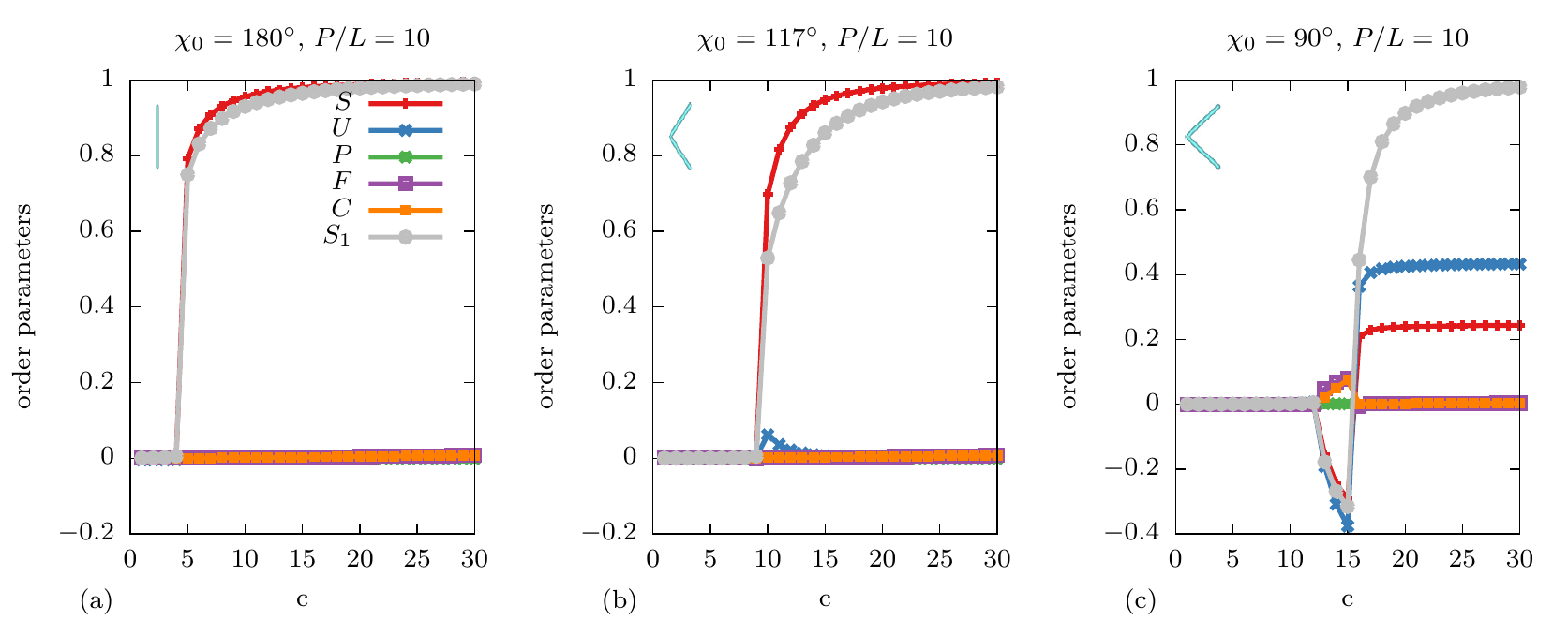}
		\caption{Order parameters as a function of density $c$ for semiflexible boomerangs ($P/L=10$) with preferred angles (a) $\chi_0=180^\circ$, (b) $\chi_0=117^\circ$, and (c) $\chi_0=90^\circ$. The key applies to (a)-(c).}\label{fig:orderPar10}
	\end{figure*}

Next, in Fig.~\ref{fig:orderPar10} we consider the order parameter trends of semiflexible boomerangs with $P/L=10$ and with preferred opening angles of (a) $\chi_0=180^\circ$, (b) $\chi_0=117^\circ$, and (c) $\chi_0=90^\circ$. In Fig.~\ref{fig:orderPar10}(a), for boomerangs with a preferred straight configuration, there is an I-N$_+$ transition as in the case of stiff boomerangs, but this has shifted to higher densities with $c_i \approx 4.05$ and $c_n \approx 4.54$. The density gap $c_n-c_i$ is therefore also reduced compared with stiffer rods, in agreement with flexible needles in the continuum limit~\cite{khokhlov1981,khokhlov1982,dijkstra1995,wessels2003,wessels2006,dennison2011JCP}. In the case of Fig.~\ref{fig:orderPar10}(b), after the isotropic-prolate nematic transition, these semiflexible boomerangs do not transition to a biaxial nematic phase as their stiff counterparts did, but instead deform from their preferred angle $\chi_0=117^\circ$ to straighter configurations in the prolate nematic phase, as also discussed above. In Fig.~\ref{fig:orderPar10}(c), the boomerangs have an I-N$_-$ transition as they did in the stiff case, but instead of forming a D$_4$ at high densities, they rather deform into straighter boomerangs and form an N$_+$ phase. However, as discussed above, the segmentwise approximation breaks down and we no longer trust our calculation at $c>15$ in Fig.~\ref{fig:orderPar10}(c).

\begin{figure*}[tbph]
	\centering
			\includegraphics[width=\textwidth]{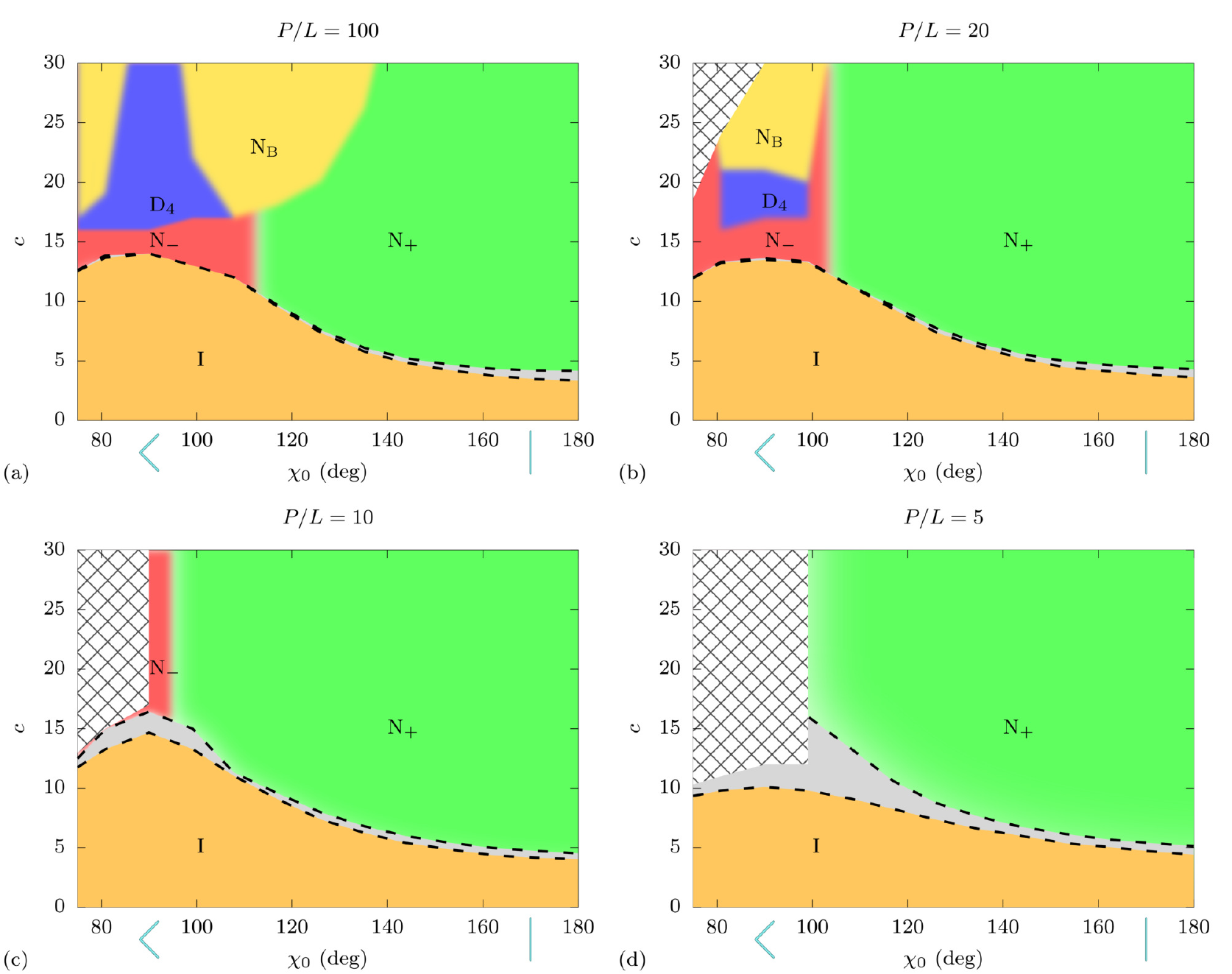}
		\caption{Phase diagrams in the preferred angle $\chi_0$ and density $c$ representation for semiflexible boomerangs with a persistence length of (a) $P/L=100$, (b) $P/L=20$, (c) $P/L=10$, and (d) $P/L=5$. Crosshatched regions denote the breakdown of the segmentwise approximation for the excluded volume. The illustrations along the horizontal axis show the particle shape for $\chi=90^\circ$ and $\chi=180^\circ$.}\label{fig:phaseDiagram100}
	\end{figure*}

We use the order parameters and the thermodynamic quantities to construct phase diagrams in the ($\chi_0$, $c$) representation in Fig.~\ref{fig:phaseDiagram100} for the four different persistence lengths: (a) $P/L=100$, (b) $P/L=20$, (c) $P/L=10$, and (d) $P/L=5$. In addition, we use the probability distribution for interarm angles $g(\chi)$ to set an approximate criterion of $\int_0^{\pi/4} d\chi \, g(\chi) > 0.1$ to signify the break down of the theory, which is shown as a crosshatched region in the phase diagrams of Fig.~\ref{fig:phaseDiagram100}. In the rigid case of Fig.~\ref{fig:phaseDiagram100}(a), we see an isotropic phase at low densities, with a transition at higher densities to a prolate nematic when $\chi_0> 112^\circ$ and to an oblate nematic when $\chi_0< 112^\circ$. This separation between prolate and oblate ordering at $\chi_0 \approx112^\circ$ is similar to the Landau angle of $\chi_0 = 107^\circ$ found for rigid boomerangs in Ref.~\cite{teixeira1998}. We do not see a direct isotropic to biaxial nematic transition due to our threshold of 0.1 for the order parameters, which is not unexpected since the order parameters are predicted to be small close to the Landau point. In addition, as discussed, we do not find an N$_\text{B}$ phase but rather a D$_4$ phase for preferred angles close to $\chi_0=90^\circ$. In Fig.~\ref{fig:phaseDiagram100}(b), we find a similar topology, but see that the flexibility destroys much of the region of biaxial nematic stability, with the prolate nematic phase encroaching on this region and the separation between the N$_-$ and N$_+$ moving to smaller angles. The mechanism is the relatively cheap energy penalty to bend the boomerangs into needle-shaped objects. For the even more flexible boomerangs in Fig.~\ref{fig:phaseDiagram100}(c), there is no longer a biaxial nematic or D$_4$ phase. Finally, in the most flexible case studied here [Fig.~\ref{fig:phaseDiagram100}(d)], we see that the region in which we predict our approximation to break down has become larger.

 In Ref.~\cite{vaghela2017}, high flexibility was shown to cause spontaneous formation of biaxial nematics from boomerangs with $\chi_0=180^\circ$, which are uniaxial on average. However, we found only uniaxial prolate nematic phases for $\chi_0=180^\circ$ even for $P/L=5$ ($\sigma_\chi \lesssim 13^\circ$) and $P/L=1$ ($\sigma_\chi \lesssim 50^\circ$) (not shown). The latter case is so flexible that even at low densities for $\chi_0=180^\circ$, $g(\chi)$ has a peak at small angles, so we no longer trust our approximation there.

\section{Discussion and Conclusions}\label{sect:conclusions}

In this paper, we used second-virial density functional theory for semiflexible chains to study the phase behaviour of hard semiflexible boomerangs with different persistence lengths and preferred angles. For stiff boomerangs, we found that the separation between prolate and oblate ordering occurs at $\chi_0 \approx 112^\circ$, which is similar to the Landau angle of $\chi_0 =107^\circ$ reported for rigid boomerangs~\cite{teixeira1998}. However, our phase diagram has a limited region of oblate nematic stability, due to the preference of platelike boomerangs to form a D$_4$ phase with four-fold rotational symmetry. This phase requires fourth-rank order parameters to identify it, and was neglected in the work of Ref.~\cite{teixeira1998} where only second-rank order parameters were considered.

 In contrast with recent results~\cite{vaghela2017}, we did not find any evidence of a biaxial nematic phase composed of flexible boomerangs with a straight preferred configuration, which are uniaxial particles on average. Moreover, we found that even for particles that are intrinsically biaxial, flexibility discourages the formation of biaxial nematic phases in favour of prolate nematic phases. The underlying mechanism that we identified here is that, at high densities, the flexible boomerangs tend to stretch out in order to reduce their excluded volume. This is similar to an experimentally observed stretching of semiflexible polymer coils in a background nematic in Ref.~\cite{dogic2004}, which was shown by theory in Ref.~\cite{dennison2011PRL}.

 Using the excluded volume in the segmentwise approximation, as was also done in other works studying boomerangs~\cite{teixeira1998,vaghela2017}, allowed us to formulate the theory in terms of single segment properties, from which the full particle orientation distribution functions and thermodynamics can nevertheless be deduced. We expect this approach to be more accurate than the method based on directly solving for the set of four second-rank order parameters as was done in Ref.~\cite{teixeira1998} for rigid boomerangs and in Ref.~\cite{vaghela2017} for flexible boomerangs. For instance, only considering the second-rank order parameters limits the possible phases that can be studied, excluding for example the D$_4$ phase. Moreover, Ref.~\cite{vaghela2017} is based on the additional approximation of interpolating the excluded volume between six known configurations in order to write it in terms of four angles (three relative Euler angles plus one interarm angle), even though for the flexible case actually five angles would have been needed within this method: three relative Euler angles plus the interarm angles of both particles. Note however that the segmentwise excluded volume in terms of the segment orientations only depends on four angles, the cosines of which being the dot products of the orientations of each pair of segments. Our method not only yields richer information as we have the full boomerang orientation distribution function, but it also has the advantage of being able to treat flexible boomerangs with a bent preferred configuration.

 However, a drawback of the currently used approach of the segmentwise approximation is that it neglects the polarity of the boomerangs, which becomes worse for very bent configurations. We saw that in the case of very flexible particles at high densities, this led to spurious results where the boomerangs tended to prefer ``closed up'' configurations with small interarm angles. The recently developed strategy to use Monte Carlo calculations to calculate the excluded volume kernel $E(\Omega,\Omega')$ more precisely~\cite{belli2014,dussi2015} could be used for going beyond the segmentwise approximation, and may reveal polar or chiral phases in the phase diagrams~\cite{greco2015,lubensky2002,bisi2008}. Direct computer simulations of boomerang systems are of course also a continued source of information and insight. For many years to come we will be able to build on the foundations of liquid-crystal simulations and theory~\cite{frenkel1984,frenkel1985,mulder1985,eppenga1984,frenkel1987,bolhuis1997} laid by Daan Frenkel.

\section*{Acknowledgments}

It is a great pleasure to congratulate Daan Frenkel on the occasion of his 70th birthday. His deep understanding and broad knowledge of science combined with his wit and enormous recollection of (non-)trivia on literally any topic have impressed RvR from his undergraduate days onward, and made every interaction with Daan a privilege and a pleasure. RvR is grateful for many years of Daan's guidance, support, and inspiration. We wish Daan good health and spirit for many years to come.

We thank Massimiliano Chiappini, Marjolein Dijkstra, and Simone Dussi for useful discussions. This work is part of the D-ITP consortium, a program of the Netherlands Organization for Scientific Research (NWO) that is funded by the Dutch Ministry of Education, Culture and Science (OCW). We also acknowledge financial support from an NWO-VICI grant.

\bibliographystyle{tfo}

\bibliography{boomerangs_ref}

\end{document}